\documentclass[10pt]{article}
\usepackage[paperheight=9in,paperwidth=6.3in,top=0.6in,bottom=1in,right=0.55in,left=0.55in]{geometry}
\usepackage{amsmath, amssymb, mathrsfs}
\newcommand{\pa} [2] {\frac{\partial #1}{\partial #2}} 

\begin{document}

\title{Obstacles to the quantization of general relativity using symplectic structures}
\author{Tom McClain \\
Department of Physics and Engineering \\
Washington and Lee University }
\date{}

\maketitle

\large \begin{center} \textbf{Abstract} \end{center} 
\normalsize In this paper I give overviews of the polysymplectic approach to covariant Hamiltonian field theory and the simplest geometric quantization of classical particle theories. I then give a synopsis of a recently proposed toy model for applying this geometric quantization map to polysymplectic field theory. I show that no special difficulties arise when this toy model is applied to general relativity. I then sketch the reasons why the standard tools of covariant Hamiltonian field theory are not up to the challenge of GR, so that the resulting quantum theory cannot be taken seriously. A few remarks are given about prospects for future work.

\section{Introduction}
The topic of covariant Hamilonian field theory goes back to the pioneering work of Dedonder \cite{de1930theorie} and Weyl \cite{weyl1935geodesic} in the 1930s, and the modern approach to the subject exists in many forms: multisymplectic \cite{kijowski1975multisymplectic}, polysymplectic \cite{gunther1987polysymplectic}, etc. But in most of these modern approaches, the subject begins with and builds most of it structures upon the basis of the Lagrangian approach to classical field theory and, in particular, the geometric structure of the jet bundle. The exception is Christian G\"unther's work, the so-called polysymplectic approach. Recently, I proposed a novel modification to the polysymplectic approach that continues to build upon its simpler geometric foundation, but is truly global in character \cite{mcclain2020global}. This approach utilizes symplectic structures much more similar to that of the standard geometric approach to covariant Hamiltonian particle theory \cite{mcclain2018some}. \\

One major advantage of this approach is that it allows for a relatively straightforward generalization of the original Kostant-Souriau quantization map from geometric quantization that applies more-or-less directly to field theories \cite{mcclain2018some}. Though this quantization map has a number of nice features \cite{mcclain2020tautological}, there is at present no direct correspondence between its output and that of ordinary quantum field theory. At present, it can therefore only be regarded as a toy model for the geometric quantization of classical field theories. \\

The goal of this paper is to show that there are nonetheless important lessons that can be learned from this particular toy model. Of special importance is the case of general relativity: since no perturbatively renormalizable quantum field theory of gravity exists, there is no ``ordinary quantum field theory" in this case and many of the lessons that can be learned must come instead from a geometric (or other non-canonical) approach to quantization. \\

In particular, I will show in section \ref{KSGR} that no special difficulty is encountered in the generalized Kostant-Souriau quantization of general relativity. Perhaps somewhat astonishingly, the basic approach I will present in section \ref{KSgeneral} applies without modification to this generally very challenging special case! However, I will discuss in section \ref{problems} that there are several purely classical obstacles that prevent us from taking the quantization procedure of section \ref{KSGR} seriously. I will briefly remark on some possible solutions to these problems as and where appropriate. \\

\section{Covariant Hamiltonian Field Theory in the Polysymplectic Approach} \label{structures}
Since our goal is to quantize general relativity on the simplest realistic space-time manifold $\mathbb{R}^4$, much of the sophisticated (and challenging) analysis of \cite{mcclain2020global} is unnecessary. Therefore, although all the analysis to come can be generalized to apply to an arbitrary differentiable manifold I will keep things as simple as possible by making explicit use of the various nice properties of $\mathbb{R}^4$, including its canonical coordinates, metric, and (flat) connection. This analysis will therefore in many ways more closely resemble the long-known polysymplectic theory of \cite{gunther1987polysymplectic} rather than my recent analysis in \cite{mcclain2020global}. \\

The analysis begins with a differentiable manifold representing space-time: $M = R^4$. Any classical field theory -- including general relativity -- will ultimately be formulated in terms of sections of a fiber bundle $E$ whose base manifold is $M = \mathbb{R}^4$. Sections of $E$ represent physical fields. For example, the theory of a single scalar-valued field would be analyzed in terms of sections of $E = R \times M$, whereas from the metric perspective general relativity would be analyzed in terms of symmetric sections of the bundle $E = T^*M \otimes T^*M$. If we assume that $E$ is a vector bundle -- as it is in the previous two examples, and most other classical field theories -- then we have global fibered coordinates on $E$ in which a point $e \in E$ can be represented as 
$$ e = x^\alpha e_\alpha + \phi^I e_I $$

The extended phase space of such a field theory is the space $P = V^* E \otimes TM$, as it is points of this space that identify the full physical state of a field in the polysymplectic approach of \cite{mcclain2020global}. (Here $VE$ is the vertical bundle of $E$, defined as the space of all $v \in TE \mid \epsilon_*(v) = 0$, where $\epsilon : E \to M$ is the projection operator associated with $E$.) This can be seen by looking at how points are represented in global fibered coordinates on $P$ compatible with the global fibered coordinates just given on $E$:
$$ p \in P = x^\alpha e_\alpha + \phi^I e_I + \pi^\alpha_I d \phi^I \otimes \pa{}{x^\alpha} $$
Intuitively, the new coordinates $\pi^\alpha_I$ represent the several momenta conjugate to each field configuration degree of freedom. Each field variable carries $n = \dim M$ conjugate momenta, as is standard in covariant formulations of Hamiltonian field theory. \\

One of the many advantages of having global fibered coordinates is that I can easily define the various symplectic structures simply by giving their representations in these special coordinates. Since they are all tensors, they are equally well-defined in any coordinate system I may wish to transform into. The first such (poly)symplectic tensor I will define is the analog of the ordinary tautological one-form. It is a section of the bundle $T^*P \otimes TM$ represented in global canonical coordinates as:
$$ \theta = x^\alpha e_\alpha + \phi^I e_I + \pi^\alpha_I d \phi^I \otimes \pa{}{x^\alpha} + 0 \ dx^\beta \otimes \pa{}{x^\alpha} + \pi^\alpha_I d \phi^I \otimes \pa{}{x^\alpha} + 0 \ d \pi^\beta_I \otimes \pa{}{x^\alpha}$$
Intuitively, at each point this vector-valued one-form has as its kernel those vectors in $TP$ that are tangent to $M$ or to the fibers of $P_e$. To show that this map can be defined in an intrinsic, coordinate-invariant way in the general case of a non-flat differentiable manifold $M$ requires more sophisticated analysis which I will avoid here. See \cite{mcclain2020global} for details. \\

The map $\theta$ -- which I call the tautological tensor in virtue of its similarities in form and function to the ordinary tautological one form of geometric Hamiltonian particle theory \cite{abraham1978foundations} -- is the first important symplectic tensor of the theory. The next is the symplectic tensor $\omega$, defined in this simplified analysis by taking the exterior derivative of the tensor field $\theta$ in global fibered coordinates, yielding the section of $T^*P \otimes T^*P \otimes TM$ given by 
$$\omega = d \theta = d \pi^\alpha_I \wedge d \phi \otimes \pa{}{x^\alpha}$$
(This too is more challenging in the case that $M$ is a more generic (and, in particular, non-flat) differentiable manifold. See \cite{mcclain2020global} for more details.) \\

The final structure I will define is the inverse of the symplectic tensor. In virtue of its similarities to the ordinary Poisson bi-vector on a Poisson manifold \cite{flato1976deformations} I call it the Poisson tensor and define it in global fibered coordinates by  
$$ \Pi = \pa{}{\phi^I} \wedge \pa{}{\pi^\alpha_I} \otimes dx^\alpha $$
It is, unsurprisingly, a section of $TP \otimes TP \otimes T^*M$. (Yet again, this map is easier to define in global coordinates than in the general case; see \cite{mcclain2020global} for details.) \\

The diligent reader will have noticed that so far I have defined a number of symplectic tensors over a rather strange looking fiber bundle, but I have yet to make any contact with classical field theory! Given a Hamiltonian function $H : P \to \mathbb{R}$ encoding the dynamics of a particular classical field and a section $\gamma : M \to P$ representing a physical field configuration, those sections that are physically realizable must obey
$$ \omega(v, T \gamma) = dH(v) $$
for all vertical vector fields $v : P \to TP \mid d (\epsilon \circ \pi)(v) =0 $ (where $\epsilon \circ \pi : P \to E \to M$ is the composition of the projection maps $\pi : P \to E$ and $\epsilon: E \to M$ and $d(\epsilon \circ \pi)$ is its differential). In the global fibered coordinates employed throughout, this intrinsic condition yields the two equations
$$\pa{\gamma^I}{x^\alpha} = \pa{H}{\pi^\alpha_I}$$
and
$$\pa{\gamma^\alpha_I}{x^\alpha} = - \pa{H}{\phi^I}$$
which are the Dedonder-Weyl equations -- the covariant form of Hamilton's field equations. They are equivalent to the Euler-Lagrange equations of motion for hyper-regular Lagrangians \cite{gotay1998momentum}.

\section{Kostant-Souriau Quantization for Field Theories} \label{KSfields}
\subsection{The Kostant-Souriau Quantization Map} \label{KSparticles}
With these structures in hand -- the polysymplectic counterparts of the tautological one-form, symplectic form, and Poisson bi-vector from ordinary symplectic geometry -- it is possible to straightforwardly generalize the Kostant-Souriau quantization map of geometric quantization so that it applies to field theories. It is worth pointing out here at the beginning that in the more general case it is only the difficulty of intrinsically defining these geometric structures that increases; actually defining the KS quantization map remains mostly unchanged. See \cite{mcclain2018some} for more discussion of this crucial point. \\

First, a brief reminder of how the Kostant-Souriau quantization map works in the particle case. Given a symplectic manifold $P$ with tautological one-form $\theta$, symplectic form $\omega$, and Poisson bi-vector $\Pi$, the Kostant-Souriau quantization map takes functions on the phase space $P$ and maps them to linear operators acting on complexified phase space functions by
\begin{equation}
Q_{KS} : C^\infty(P, \mathbb{R}) \to \text{Lin}(C^\infty(P, \mathbb{C}), C^\infty(P, \mathbb{C})) \mid f \mapsto f + \theta(X_f) + \text{i} \hbar X_f
\end{equation}
where $X_f$ is the Hamiltonian vector field associated with the function $f$ via either
$$ \omega(X_f, -) = df  $$
or
$$ X_f = \Pi(df, -) $$

In local canonical coordinates $\{q^i, p_i \}$ on $P$ in which the tautological one-form is given by $\theta = p_i dq^i$ (the ordinary KS quantization map requires no assumptions about the flatness of the base space), the map is given by
$$ Q_{KS} \mid f \mapsto f - p_i \pa{f}{p_i} + \text{i} \hbar \pa{f}{q^i} \pa{}{p_i} - \text{i} \hbar \pa{f}{p_i} \pa{}{q^i} $$
Note that, while relatively straightforward to define, this map has a number of nice features. For example, it maps the canonical coordinate functions to (mostly) appropriate looking operators:
\begin{equation}
Q_{KS}(q^i) = q^i + \text{i} \hbar \pa{}{p_i}
\end{equation}
(but note the strange looking momentum coordinate derivative) and
\begin{equation}
Q_{KS}(p_i) = - \text{i} \hbar \pa{}{q^i}
\end{equation}
It even maps the angular momentum function $L_3$ (as well as the other two) to a (mostly) appropriate looking operator:
\begin{equation}
Q_{KS}(L^3) = Q_{KS}(q^1 p_2 - q^2 p_1) = \text{i} \hbar \big( p_2 \pa{}{p_1} - p_1 \pa{}{p_2} - q^1 \pa{}{q^2} + q^2 \pa{}{q^1} \big)
\end{equation}
The presence of the momentum coordinate derivatives in these operators is embarrassing, but if these could be eliminated then the operators look pretty good. However, it is easy to see that this map is far from perfect. For example, the one dimensional simple harmonic oscillator Hamiltonian maps to 
\begin{equation}
Q_{KS}(H_{SHO}) = Q_{KS} (\frac{p^2}{2m} + \frac{1}{2} m \omega^2 q^2) = \text{i} \hbar (- \frac{p}{m} \pa{}{q} + m \omega^2 q \pa{}{p}) - \frac{p^2}{2m} + \frac{1}{2} m \omega^2 q^2
\end{equation}
which is a far cry from the expected
\begin{equation}
Q(H_{SHO}) = - \hbar^2 \pa{^2}{q^2} + \frac{1}{2} m \omega^2 q^2
\end{equation}
These problems can be mostly eliminated, either by the sophisticated application of geometric quantization techniques like polarization and metaplectic correction \cite{woodhouse1997geometric}, or by the simpler application of tautological tuning \cite{mcclain2020tautological}. Since I am only trying to build a toy model I will not take either of these paths here. See \cite{mcclain2018some} for a preliminary analysis of how the techniques of \cite{mcclain2020tautological} can be applied to the case of quantizing field theories.

\subsection{The General Theory} \label{KSgeneral}
Given our generalizations from section \ref{structures} of the symplectic structures used in section \ref{KSparticles}, it seems intuitively reasonable that we can create a variant of the Kostant-Souriau quantization procedure simply by contracting our symplectic tensors with a vector field $v : M \to TM$ or one-form $\beta : M \to T^*M$ (as appropriate) and then applying them as usual. This procedure gives us the three reduced tensors:
$$\theta(-,\beta) = \beta_\alpha \pi^\alpha_I d \phi^I $$
$$\omega(-,-,\beta) = \beta_\alpha d \pi^\alpha_I \wedge d \phi^I $$
$$ \Pi(-, -, v) = v^\alpha \pa{}{\phi^I} \wedge \pa{}{\pi^\alpha_I} $$
Note that the reduced tautological tensor is now in fact a one-form over $P$, the reduced symplectic tensor is now a true two-form, and the reduced Poisson tensor is now a true bi-vector. The reduced tensors are naturally dependent upon the choice of vector field $v$ and one-form $\beta$. \\

This procedure can be made to work, but with two important caveats. First, it is no longer an option to define the vector fields $X_f$ via $\omega$, as $\beta(\omega)$ is in general no longer non-degenerate so that there is no guarantee that these vector fields exist. The procedure of defining $X_f$ via $\Pi$ is still viable (though it will lose some of its nice properties). The second caveat is that since it is necessary to use both $\theta$ and $\Pi$, it would be nice to find a relation between the one-form $\beta$ and the vector field $v$ to reduce the number of new structures that are introduced. Given our access to global fibered coordinates and a canonical (flat) metric on $\mathbb{R}^4$, the natural way to do this is to choose $v$ and $\beta$ to be the coordinate vector field and basis one-form corresponding to a single coordinate function, $x^\alpha$. That is, to take $v = \pa{}{x^\alpha}$, $\beta = dx^\alpha$. With these caveats, the appropriate generalization of the Kostant-Souriau quantization map is:
\begin{equation} \label{ksfieldquantization}
Q_{KS} \mid f, x^\alpha \mapsto f + \theta(X_f, dx^\alpha) + \text{i} \hbar X_f
\end{equation}
where $X_f := \Pi(df, - , \pa{}{x^\alpha})$. \\

As advertised, this map relies a great deal on the particular space-time coordinate $x^\alpha$ chosen to perform the quantization. (There are a number of other ways to define this quantization procedure that make it seem much more covariant. However, in addition to adding a good deal of complexity and making the result look much less like ordinary KS quantization, they seem also to reproduce even fewer of the results of canonical quantum field theory.) This being the case, it is essential to decide which coordinate function one should choose in order to get the ``right" quantization map. The answer that would immediately occur to anyone familiar with canonical quantum field theory is that we should choose $x^0$. With this choice, one finds
\begin{equation} \label{kszeroquantization}
Q^0_{KS} \mid f \mapsto f + \theta(X^0_f, dx^0) + \text{i} \hbar X^0_f
\end{equation}
where $X^0_f := \Pi(df, - , \pa{}{x^0})$. 
In global fibered coordinates, this looks like
$$ Q^0_{KS} \mid f \mapsto f - \pi^0_I \pa{f}{\pi^0_I} + \text{i} \hbar \pa{f}{\phi^I} \pa{}{\pi^0_I} - \text{i} \hbar \pa{f}{\pi^0_I} \pa{}{\phi^I}$$
Though only a toy model, this map -- like the Kostant-Souriau quantization map for Hamiltonian particle systems -- has a number of nice features. It too associates mostly nice looking operators with the canonical coordinates functions $\{x^\alpha, \phi^I, \pi^\alpha_I\}$:
$$Q^0_{KS} (x^\alpha) = x^\alpha$$
$$ Q^0_{KS} (\phi^I) = \phi^I + \text{i} \hbar \pa{}{\pi^0_I} $$
$$ Q^0_{KS} (\pi^0_I) = - \text{i} \hbar \pa{}{\phi^I} $$
$$ Q^0_{KS} (\pi^i_I) = \pi^i_I  $$
(here $i \in \{ 1, 2, 3 \}$, whereas $\alpha \in \{ 0, 1, 2, 3 \}$)

However, the resulting commutation relations are not the canonical ones. The only non-zero commutators are:
$$ [ Q^0_{KS}(\phi^I) , Q^0_{KS}(\pi^0_J) ] = \text{i} \hbar \delta^I_J \neq \text{i} \hbar \delta^I_J \delta(\vec x - \vec y) = [ \hat \phi^I (\vec x) , \hat \pi^0_J (\vec y) ] $$
(The hats on the operators serve to distinguish the canonical quantum field theory operators from those coming from KS quantization.) This result is entirely expected, since it is the coordinate functions on $P$ rather than sections of $P$ that are being quantized with the generalized Kostant-Souriau maps of \eqref{ksfieldquantization} and \eqref{kszeroquantization}. However, there is a close and natural link between the KS commutators and those of canonical quantum field theory:
$$ \int [ \hat \phi^I (\vec x) , \hat \pi^0_J (\vec y) ] \star dx^0 = [ Q^0_{KS}(\phi^I) , Q^0_{KS}(\pi^0_J) ] $$
where $\star$ is the Hodge star operator $\star : \Lambda^k (M) \to \Lambda^{n-k}(M)$. (In the global fibered coordinates I have been using, the assumption of the flat metric used in canonical quantum field theory gives $\star dx^0 = dx^1 \wedge dx^2 \wedge dx^3 $.) \\

Much of the discussion of generalized Kostant-Souriau quantization maps for particle systems in \cite{mcclain2020tautological} can be applied with minimal modification to the field theoretic framework discussed in this section. However, since this paper is about obstacles to the quantization of general relativity rather than successes, it makes sense to move on the particular case of GR instead of discussing these extensions. Please see \cite{mcclain2018some} for a few more details. \\

\subsection{Application to General Relativity} \label{KSGR}
Perhaps somewhat surprisingly, all of the results of section \ref{KSgeneral} can be applied to the case of general relativity with minimal modification. \\

I begin with the ordinary geometric interpretation of general relativity, in which the metric on $TM$ with $M = \mathbb{R}^4$ is taken as the fundamental field. This choice amounts, in the language of section \ref{structures}, to 
$$E = T^*M \otimes T^*M$$
which in turn leads to 
$$ P = V^*E \otimes TM $$
with global fibered coordinates on $P$ given by $\{ x^\alpha, g_{\alpha \beta}, \pi^{\alpha \beta \gamma}\}$ and a point in the phase space $p \in P$ represented by $p = x^\alpha e_a + g_{\alpha \beta} dx^\alpha \otimes dx^\beta + \pi^{\alpha \beta \gamma} d g_{\alpha \beta} \otimes \pa{}{x^\gamma} $.
(Other geometric frameworks are certainly possible, and lead to other, similar results. See \cite{mcclain2018some} for more details.) \\

Since it is still the case that $M = \mathbb{R}^4$, the fact that the physical metric is no longer flat does not obstruct these constructions: one still has access to the canonical flat metric and connection, even though these are not the standard tools of general relativity. All the symplectic structures of section \ref{structures} can therefore be imported immediately:
\begin{equation} \theta = \pi^{\alpha \beta \gamma} d g_{\alpha \beta} \otimes \pa{}{x^\gamma} \end{equation}
\begin{equation} \omega = d \pi^{\alpha \beta \gamma} \wedge d g_{\alpha \beta} \otimes \pa{}{x^\gamma} \label{omegaGR} \end{equation}
\begin{equation} \Pi = \pa{}{ \pi^{\alpha \beta \gamma} } \wedge \pa{}{ g_{\alpha \beta} } \otimes d x^\gamma \end{equation}

In this case it seems more in keeping with the physical intuitions of general relativity to choose a global time-like Killing vector field $K$ with which to form the quantization map, as the global canonical ``time" coordinate $x^0$ is not the physical time of any local observer (even one at infinity, as this would amount to a flip of sign, at least in the ``mostly-plus" signature convention). Since the presence of the metric gives us $K^* = g(K)$, the construction even goes through without additional input. In the case the $g = \eta$ (the Minkowski metric), this choice is almost identical to the one made in the construction of \eqref{kszeroquantization}. However, in keeping with the spirit of the rest of the analysis I will make the (probably erroneous) choice to stick with the $x^0$ convention of \eqref{kszeroquantization}. \\

Taking this very mathematical attitude toward the global coordinates on $M$ in identifying the appropriate vector field to use in forming the generalized Kostant-Souriau quantization map of \eqref{ksfieldquantization} gives the global fibered coordinate description (or, more physically, the description in local canonical coordinates in which $\omega$ has the same form as in \eqref{omegaGR})
$$ Q_{KS}^{0} \mid f \mapsto f - \pi^{\alpha \beta 0} \pa{f}{\pi^{\alpha \beta 0}} + \text{i} \hbar \pa{f}{g_{\alpha \beta}} \pa{}{\pi^{\alpha \beta 0}} - \text{i} \hbar \pa{f}{\pi^{\alpha \beta 0}} \pa{}{g_{\alpha \beta}} $$
This map associates the following operators with the canonical coordinate functions $\{x^\alpha, g_{\alpha \beta}, \pi^{\alpha \beta \gamma} \}$:
$$Q^0_{KS} (x^\alpha) = x^\alpha$$
$$ Q^0_{KS} ( g_{\alpha \beta} ) =g_{\alpha \beta} + \text{i} \hbar \pa{}{\pi^{\alpha \beta 0}} $$
$$ Q^0_{KS} (\pi^{\alpha \beta 0}) = - \text{i} \hbar \pa{}{g_{\alpha \beta}} $$
$$ Q^0_{KS} (\pi^{\alpha \beta i}) = \pi^{\alpha \beta i}  $$
(where again Latin letter indices take on the values $ \{ 1, 2, 3 \}$ and Greek letter indices take on the values $\{ 0, 1, 2, 3 \}$)

Again, the resulting commutation relations are not the canonical ones (though in this case we should have much less confidence in the canonical commutation relations, as they come from a process known to give perturbatively non-renormalizable results). The non-zero commutators are:
$$ [ Q^0_{KS}(g_{\alpha \beta}) , Q^0_{KS}(\pi^{\gamma \delta 0}) ] = \text{i} \hbar \delta^\alpha_\gamma \delta^\beta_\delta \neq \text{i} \hbar \delta^k_i \delta^l_j \delta(x - y) = [ \hat g_{i j} (x) , \hat \pi^{kl} (y) ] $$
It is worth noting that there are more non-zero commutators coming from the generalized KS procedure than there are from the canonical one, as the KS procedure does not single out the spatial indices for special treatment. However, as the point of this analysis is not to produce a correct quantum theory of gravity I will not debate the merits and demerits of this approach versus the canonical one. Suffice it to say that they give different results. \\

What is more important for the present analysis is that -- the merits or demerits of the generalized KS approach aside -- there was no mathematical obstacle to performing exactly the same quantization procedure for general relativity that was used for scalar fields! 

\section{Issues with the Kostant-Souriau Quantization of General Relativity} \label{problems}
The fact that no mathematical obstacle was encountered in applying the methods of generalized KS quantization to general relativity does not mean that everything is okay. It just means that the problems involved are not quantum problems! \\

The first and most serious problem is that the Legendre transformation fails in the case of general relativity. Most covariant Hamiltonian analysis suggests that we need to find the partial derivative $\pa{\mathscr{L}}{\pa{\phi^I}{x^\mu}}$ in order to define $\pi^\mu_I = \pa{\mathscr{L}}{\pa{\phi^I}{x^\mu}}$ and the covariant Hamiltonian $\mathscr{H} = \pi^\mu_I  \pa{\phi^I}{x^\mu} - \mathscr{L}$. However, in the case of general relativity we get \cite{mcclain2020global}

\begin{multline}
\pi^{\chi \psi \omega} = \pa{\mathscr{L}}{\pa{g_{\chi \psi }}{x^\omega}} = \frac{c^4}{32 \pi G} \sqrt{- \det g} \ \pa{g_{\alpha \beta }}{x^\gamma} \times \\
\bigg(
- 4 g^{\alpha \psi } g^{\beta \gamma} g^{\chi \omega} + 2 g^{\alpha \chi } g^{\beta \psi } g^{\gamma \omega} - 2 g^{\alpha \beta } g^{\gamma \omega} g^{\chi \psi } + g^{\alpha \beta } g^{\gamma \psi } g^{\chi \omega} + g^{\alpha \gamma} g^{\beta \omega} g^{\chi \psi } \bigg)
\end{multline}

Even if this were invertible to solve for the $\pa{g_{\alpha \beta}}{x^\gamma}$ in terms of the $\pi^{\chi \psi \omega}$, $R$ (and therefore the Lagrangian density $\mathscr{L}$) is a function of the second derivatives $\pa{^2 g_{\alpha \beta}}{x^\gamma \partial x^\delta}$, and these are certainly not encoded in the $\pi^{\chi \psi \omega}$. So the Legendre transformation fails. We therefore do not have a well-defined Hamiltonian function, and therefore no Hamilton's equations to define the classical theory. The analysis of the \ref{KSGR} is therefore not to be trusted: if the classical theory is not well defined, then it makes no sense to quantize it! \\

This problem is by no means restricted to the polysymplectic approach, afflicting all covariant Hamiltonian field theories that begin with the same GR geometry. Many efforts have therefore been made to fix it, including higher order Legendre transformations \cite{magnano1990legendre}, alternative geometric formulations \cite{mcclain2018some}, and more. None of the alternative geometric formulations seems to fix the problem, and it is not clear that higher order Legendre transformations are compatible with the polysymplectic approach. It may be possible, however, to define an appropriate Hamiltonian directly; work in this area is ongoing. \\

A second, independent problem with this approach is that the global time coordinate singled out in \ref{KSGR} need not be the local time coordinate of any physical observer. This is a physical, rather than a mathematical problem. It is possible to remove this problem by choosing a global time-like Killing vector field and its metric dual in place of the global time coordinate. However, the result is even farther from the canonical one.

\section{Conclusions} \label{conclusions}

It seems rather amazing that there are no geometric obstacles to a toy model geometric quantization of general relativity using an extended Kostant-Souriau quantization map. Perhaps slightly less amazing is that the well known and rather severe obstacles to finding a simple and consistent covariant Hamiltonian approach to general relativity make this quantization more-or-less meaningless. These obstacles must be eliminated if any project like the toy model quantization of \ref{KSGR} is to succeed. \\

\bibliographystyle{unsrt}
\bibliography{MMQGR.bib}

\end{document}